\documentclass[12pt]{article}
\usepackage{amsbsy}
\usepackage{amsfonts}
\usepackage{amssymb}
\usepackage{amsmath}
\usepackage{graphicx}
\usepackage{subfigure}
\usepackage[usenames, dvipsnames]{color}

\makeatletter
\renewcommand\section{\@startsection {section}{1}{\z@}%
                                   {-3.5ex \@plus -1ex \@minus -.2ex}%
                                   {2.3ex \@plus.2ex}%
                                   {\normalfont\large\bfseries}}
\renewcommand\subsection{\@startsection{subsection}{2}{\z@}%
                                   {-3.25ex\@plus -1ex \@minus -.2ex}%
                                   {1.5ex \@plus .2ex}%
                                   {\normalfont\normalsize\bfseries}}
\makeatother

\def\be{\begin{eqnarray}}
\def\ee{\end{eqnarray}}
\def\Tr{{\rm Tr}}

\def\NeqFour{{\cal N}=4}

\begin{document}

\vspace{ -3cm}
\thispagestyle{empty}
\vspace{-1cm}

\rightline{}

\begin{center}
\vspace{1cm}
{\Large\bf

On correlation functions of Wilson loops, local and non-local operators

\vspace{1.2cm}

   }

\vspace{.2cm}
 {Oluf~Tang~Engelund\footnote{ote5003@psu.edu}
and
   Radu~Roiban\footnote{radu@phys.psu.edu}}
\\

\vskip 0.6cm

{\em

Department of Physics, The Pennsylvania  State University,\\
University Park, PA 16802, USA

 }

\vspace{.2cm}

\end{center}

\begin{abstract}

We discuss and extend recent conjectures relating partial null limits of correlation
functions of local gauge invariant operators and the expectation value of null
polygonal Wilson loops and local gauge invariant operators. We point out that
a particular partial null limit provides a strategy for the calculation of the anomalous 
dimension of short twist-two operators at weak and strong coupling.

\end{abstract}

\newpage

\section{Introduction and discussion\label{intro}}

Increasingly efficient perturbative computational techniques for higher-order calculations
and increasingly efficient use of integrability of the worldsheet theory in $AdS_5\times S^5$
have brought about, in the last few years, remarkable progress in our understanding of
${\cal N}=4$ super-Yang-Mills (sYM) theory 
and have exposed unexpected and fascinating relations between {\it a priori} 
unrelated quantities.
Such a connection, proposed in \cite{AM}, links gluon scattering amplitudes with
the expectation value of certain null polygonal Wilson loops in ${\cal N}=4$ sYM theory.
Initially suggested at strong coupling based on the AdS/CFT correspondence, this relation
was generalized and successfully tested at weak coupling \cite{AmpWLweak} as well.

Correlation functions of local gauge invariant operators are natural observables in a conformal
field theory such as $\NeqFour$ sYM theory. Two-point functions are determined by the dimension
of the operators; since the latter are known, at least in principle, through use of integrability,
so are  the two-point functions. Apart from the dimension and $SO(3,1)\times SO(6)$ charges
of operators, three point functions are determined by additional coupling-constant dependent
"structure functions" whose evaluation is less clear. It is possible to argue that, if some of the
operators carry large quantum numbers, the calculation can be carried out in a semiclassical
expansion \cite{3PFsemiclassics}.

Even though the axioms of conformal field theories guarantee that higher-point correlation
functions are determined  by the  two- and three-point functions, an explicit evaluation along
this line is not  straightforward. It is therefore interesting to devise methods to directly
evaluate them either for generic position of operators or in special limits.

It was recently suggested \cite{AEKMS} that the duality between Wilson loops and  scattering
amplitudes can be extended to include certain special classes of correlation functions. More
precisely, for operators in the stress tensor multiplet, the following relation should hold in the planar limit:
\be
\lim_{x_{i,i+1}^2\to 0}
\frac{\langle {\cal O}(x_1)\dots {\cal O}(x_n)\rangle}{\langle {\cal O}(x_1)\dots {\cal O}(x_n)\rangle_0}
=
\frac{\langle W_n\rangle^2_{\text{fund}}}{\langle W_n\rangle^2_{0,\text{fund}}}
=
\frac{A^2(k_i=x_i-x_{i-1}, i=1,\dots, n)}{A^2_{\text{tree}}(k_i=x_i-x_{i-1}, i=1,\dots, n)}
\label{allKS}
\ee
where $W_n$ is a null polygonal Wilson loop with corners at positions $x_i$ with
$i=1,\dots,n$. Initially proposed as a relation between correlation functions of bosonic
operators, bosonic Wilson loops and MHV amplitudes, this triality conjecture was extended to
supersymmetric correlators, supersymmetric Wilson loops and generic superamplitudes
\cite{Eden:2011yp, Eden:2011ku}.
This relation was proven at the level of the unregularized integrand in
\cite{ABMSproof, Bullimore:2011ni} and shown to hold in explicit examples
\cite{Eden:2011yp, Eden:2011ku} in the presence of a dimensional regulator.

It was moreover suggested and explicitly demonstrated for  four-point correlation
functions though two-loop order \cite{EKS} that the relation (\ref{allKS})
is not restricted to operators in the tensor multiplet, but rather holds for more general
1/2-BPS operators, including operators with large quantum numbers.

Another class of interesting observables is provided by the correlation function of Wilson
loops and local operators.\footnote{For BPS  (circular)  Wilson loops  and their
generalizations such correlators have
been studied in \cite{cor,za02,zp,gom,alt}, see also \cite{miwa}. As pointed out in \cite{alt},
since null polygonal loops are, in a sense, ``locally-BPS'', they may be considered as
natural generalizations of circular loops. Since they are closed under conformal
transformations, conformal invariance restricts the form of such correlation
functions \cite{ABT}.}
Such quantities are interesting for several reasons. For example, they characterize the
expansion of a Wilson loop in local operators
\be
\frac{W(\gamma)}{\langle W(\gamma)\rangle} = \sum_i\;c_i(\gamma)\, O_i(x)~;
\ee
the coefficients $c_i$ may be found in the obvious way in terms of the correlation function
$\langle W {\cal O}_j(y)\rangle$ and the two-point function
$\langle {\cal O}_i(x) {\cal O}_j(y))$. Moreover, they may be used to factorize the expectation
value of a product of two Wilson loops or, if the coefficients $c_i$ are known, to simply evaluate
this expectation value:
\be
\frac{\langle W(\gamma)W(\gamma')\rangle}
{\langle W(\gamma)\rangle\langle W(\gamma')\rangle}
&=&
\sum_{i,j}\;c_i(\gamma)c_j(\gamma')\, \langle{\cal O}_i(x)\,{\cal O}_j(y)\rangle
\cr
&=&\sum_{i}\;c_i(\gamma) \langle{\cal O}_i(x)\, W(\gamma')\rangle
\cr
&=&\sum_{i,j}\;{\cal F}_{ij}(x-y)\langle W(\gamma)\,{\cal O}_i(x) \rangle
                                                     \langle{\cal O}_j(y)\, W(\gamma')\rangle \ ,
\label{factorization}
\ee
where ${\cal F}_{ij}(x-y)$ are the relevant conformal blocks.

A second motivation for analyzing correlation functions of Wilson loops and local operators
is that, for a special choice of operator (given by the chiral Lagrangian), they
contain information about the higher-loop corrections to the expectation value of the Wilson
loop. This is akin to the Lagrangian insertion formalism \cite{Eden:2000mv}
used in \cite{EKS} to evaluate
loop corrections to correlation functions of local operators in the null separation limit.

A generalization of the relation between correlation functions of operators and null polygonal
Wilson loops was recently proposed in \cite{ABT}. More precisely, starting with an
$(n+1)$-point correlation function and taking the limit in which $n$ points are sequentially
null-separated one should find that
\be
\lim_{x_{ii+1}^2\to 0}
\frac{\langle {\cal O}(x_1)\cdots {\cal O}(x_n){\cal O}(a)\rangle}
{\langle {\cal O}(x_1)\cdots {\cal O}(x_n)\rangle}
\sim
\frac{\langle W_n\,{\cal O}(a)\rangle}
{\langle W_n\rangle}
\label{ABT}
\ee
where the expectation values on the right-hand side are taken in the fundamental representation.
Non-vanishing values for both correlators,
${\langle {\cal O}(x_1)\cdots {\cal O}(x_n)\rangle}$
and
${\langle {\cal O}(x_1)\cdots {\cal O}(x_n){\cal O}(a)\rangle}$,
require that the total R-charges of the product of operators in the
two correlators are zero. It therefore follows that the operator
${\cal O}(a)$ must have vanishing R-charges.\footnote{Indeed the operator
discussed in detail in \cite{ABT} -- the operator dual to the string dilaton --
obeys this condition.}${}^,$\footnote{It is interesting to note that, integrating over the position of the operator ${\cal O}(a)$, the correlation function acquires the interpretation of form factor
at zero momentum. Indeed, it was argued in \cite{Alday:2007he, Maldacena:2010kp} that
form factors may be interpreted in terms of the expectation value of certain zig-zag
Wilson loops. If the position of the operator is integrated over, {\it i.e.} if the momentum
inflow though it vanishes, the zigzag Wilson line becomes closed.}

Generalizations of (\ref{ABT}) are fairly straightforward to formulate. For example,
\be
\lim_{x_{ii+1}^2\to 0}
\frac{\langle {\cal O}(x_1)\cdots {\cal O}(x_n){\cal O}(a_1)\dots {\cal O}(a_m)\rangle^{\text{conn}}}
{\langle {\cal O}(x_1)\cdots {\cal O}(x_n)\rangle}
\sim
\frac{\langle W_n\,{\cal O}(a_1)\dots {\cal O}(a_m)\rangle^{\text{conn}}}
{\langle W_n\rangle}
\label{WW}
\ee
where the upper index denotes a restriction to the connected part of the correlation functions and the points $a_j$ with $j=1,\dots,m$ are generic. One may moreover consider a second limit in which the points  in this set become sequentially null separated while remaining at generic positions with respect to the points $x_i$, $i=1,\dots,n$. In this limit the original correlation function equals the correlation function of two null polygonal Wilson loops.

Ratios of the type (\ref{ABT}), (\ref{WW}) are good observables. Indeed, as discussed in detail in \cite{EKS, AEKMS}, in the null limit correlators develop the same type of singularities as null polygonal Wilson loops. Since these divergences are located around the Wilson loop cusps or, alternatively, around the operators that are sequentially null separated, in the ratios (\ref{ABT}) and (\ref{WW}) these divergences cancel out leaving behind a finite quantity, which should exhibit the symmetries of the theory. In particular, the conformal symmetry of the correlation functions should be realized. It would be interesting to understand whether the additional
operator insertions in eqs.~(\ref{ABT}) and (\ref{WW}) have any interpretation in terms
of scattering amplitudes when the operators' position is not integrated over.

Here we will prove, to all orders in a weak coupling expansion and in the regularized
theory\footnote{We will use dimensional
regularization with $d=4-2\epsilon$ with $\epsilon>0$ and take the null separation limit
for generic $\epsilon$. Since we will keep the complete $\epsilon$ dependence, our arguments
hold in all dimensions.},
the first equality in equation~(\ref{allKS}) as well as equations (\ref{ABT}) and (\ref{WW}) for
all twist-2 operators and for a  finite rank of an $SU(N)$ gauge group.\footnote{For finite-rank
gauge groups $\langle W_n\rangle_{\text{fund}}^2$ is replaced with $\langle W_n
\rangle_{\text{adj}}$, as mentioned in \cite{EKS}.}
We will also discuss two-field operators outside this class, containing fermions and field strengths.
While not identical in details, our strategy will be similar in spirit with the eikonal line
arguments used in \cite{AEKMS, EKS}; we will separate the Feynman diagrams contributing
to correlation functions into classes depending on whether  or not there exists R-charge flow between
operators and show that certain sequences of propagators between null-separated points
are equivalent to null Wilson lines.
We will also show that diagrams in which there is no R-charge  exchange between charged
fields at null-separated points have softer singularities  than if R-charge flow is present.
We will then extend these arguments to larger classes of
operators and also to larger
classes of correlation functions, in which not all points are sequentially null-separated; we will refer to them as "partial null limits".  We will also see that
the relation between correlation functions in the null separation limit and the expectation
value of null polygonal Wilson loops is
not restricted to four-dimensional gauge theories but rather holds in all dimensions.

The arguments in this note point to a generalizations of (\ref{ABT}) and (\ref{WW}) to limits in which any operator is null-separated from at least one and at most two other operators. 
In this limit a correlation function reduces to the correlation function of certain non-local 
operators built out of fundamental fields and open Wilson lines. Let us consider the correlator 
of  two $\Tr[Z^2]$ and two $\Tr[{\bar Z}^2]$ operators at positions $x_{1},\dots,x_{4}$ in 
the limit $x_{12}^2=0=x_{34}^2$ and all other distances being nonzero. Denoting by
\be
W(x, y)={\rm P}e^{\int_x^{y} A} \ ,
\ee
it is not difficult to see that 
\be
\lim_{\stackrel{\scriptstyle{x_{12}^2\rightarrow 0}}{x_{34}^2\rightarrow 0}}
\langle {\cal O}(x_1){\cal O}(x_2){\cal O}(x_3){\cal O}(x_4)\rangle \propto
\langle \Tr[Z(x_1)W(x_1, x_2){\bar Z}(x_2)]\Tr[Z(x_3)W(x_3, x_4){\bar Z}(x_4)]\rangle \ .
\label{corropenWL}
\ee
Clearly, the open Wilson lines are null.
The proportionality coefficient depends on the regularization scheme and on the order of limits.
This generalization provides a direct link between four-point correlators of BPS operators and 
two-point functions of non-BPS operators. Indeed, each non-local operators
$\Tr[Z(x)W(x, y){\bar Z}(y)]$ may be expanded in twist-two operators\footnote{
The expansion is
\be
\Tr[Z(x)W(x, y){\bar Z}(y)]\big|_{(x-y)^2=0} = \sum_{n=0}^\infty \frac{1}{n!} 
\, (x-y)^{\mu_1}\dots (x-y)^{\mu_n} \, \Tr[Z(x)D_{\mu_1}\dots D_{\mu_n} {\bar Z}(x)]
\ .
\ee
}
thus reducing the four-point correlator reduces to a superposition of two-point functions.
Making explicit this decomposition should allow one to read off the anomalous dimensions 
of twist-two operators. This approach may be particularly efficient for the lowest twist-two 
operators -- $\Tr[Z{\bar Z}]$ -- which is a member of the Konishi multiplet and may offer 
an alternative approach to the calculation of anomalous dimensions of short operators at 
strong coupling.

\

The rest of this note is organized as follows. In \S~\ref{generalities} we discuss general
features of correlation functions, define the regularization scheme and the null limit, and
outline the proof of relations~(\ref{allKS}), (\ref{ABT}) and (\ref{WW}). Later sections
contain some of the details completing this proof.
We proceed in \S~\ref{no_insertions} to discuss the correlation function
of 2-field scalar operators in the stress tensor multiplet in the null separation limit and identify
the relevant Feynman diagrams that contribute in this limit.
We then proceed in \S~\ref{w_insertion}
to extend the discussion in \S~\ref{no_insertions} to the case of $(n+1)$-point correlation function
with $n$ null-separated points. We will finish this section with comments on the more general
correlators of the type mentioned in equation~(\ref{WW}).  In \S~\ref{large_charge} we discuss
twist-2 operators with higher spin and extend the results described in the previous sections to
their correlation functions. In \S~\ref{comments} we comment on other weak
and  strong coupling features of the correlator/Wilson loop relations.
Appendix~\ref{opotherfields} details the null and partial-null
separation limit of correlation functions
of two-field operators constructed from fermions and gauge fields.


\section{Correlation functions in null and partial null limits \label{generalities}}

Let us consider the correlation function of some number of operators of length~2 in a gauge
theory with an $SU(N)$ gauge group; we will keep $N$ arbitrary and discuss the large $N$
limit at the end.
Such a correlator is symmetric under the permutation of positions of identical operators; for
example, the correlation function of $n$ operators in the ${\bf 20}'$ representation of $SO(6)$
will be symmetric under the permutation of positions of all operators. Taking the limit in which
the operators are null-separated requires choosing a specific sequence of positions,
{\it e.g.}  $x_1, \dots, x_n$ and setting $|x_{i, i+1}|^2\rightarrow 0$.
This choice breaks their permutation symmetry to one of its cyclic
subgroups~\footnote{This limit also breaks the permutation symmetry of integrands of
higher-loop four-point correlation functions recently identified  in \cite{Eden:2011we}.};
different choices of sequences will lead, in this limit, to different dominant terms which,
through the correlator/amplitude relation \cite{EKS}, are related to (squares of) different
color-ordered amplitudes.

\begin{figure}[ht]
\begin{center}
\includegraphics[height=45mm]{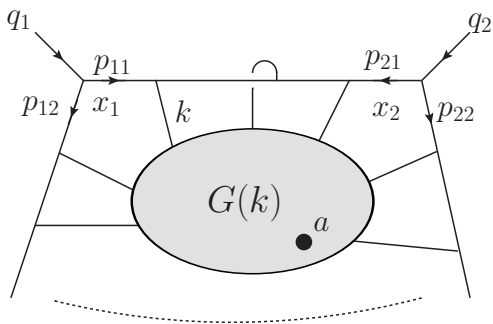}
\caption{ Presentation of a generic correlation function in terms of a Green's
function and sequences of propagators between operator insertion points which
are taken to be null separated. Additional operators, which do not participate in
this limit and are denoted by a heavy dot, may also be present. \label{general}}
\end{center}
\end{figure}

To take the limit in which (some) operators are null-separated it is useful to start with the momentum space correlation function and Fourier-transform it to position space.
In momentum space each operator ${\cal O}_i$ carries nontrivial momentum ---\\
${\widetilde{\cal O}}(q)=\int d^d x \exp(-iq\cdot x){\cal O}(x)$ ---
which is split between the fields composing it, $q_i = p_{i1}+p_{i2}$, as shown in
fig.~\ref{general}.
One may arbitrarily choose a sequence of propagators connecting adjacent points;
denoting this sequence by ${\tilde L}(p_{ij}, p_{i+1,j},k)_{A}$ with $k$ the momenta
of the lines attaching this sequence to the rest of the diagram and $A$ the corresponding
indices, and by $G(k)_{A_1\dots A_n}$ the Green's function\footnote{This Green's
function contains both connected and disconnected components.} obtained by removing
all ${\tilde L}(p_{ij}, p_{i+1,j},k)_A$, it is easy to see that
\be
\langle {\widetilde{\cal O}}_1(q_1)\cdots {\widetilde{\cal O}}_n(q_n)\rangle =
\int \prod_{i=1}^n\prod_{j=1}^2d^dp_{ij}[d^dk]\delta^d(q_i-p_{i1}-p_{i2})
{\widetilde L}(p_{ij}, p_{i+1,j},k)_{A_i} \; G(k)_{A_1\dots A_n} \ ,
\ee
where the measure factor $[d^dk]$ stands for integration over the momenta of all external lines of $G(k)$.
The factors ${\tilde L}$ may contain momentum factors arising from derivatives present in operators.
Fourier-transforming back to position space implies that
\be
\langle {{\cal O}}_1(x_1)\cdots {{\cal O}}_n(x_n)\rangle =
\int [d^dk] \prod_{i=1}^n \prod_{j=1}^2\int d^d p_{ij}\,
{\overrightarrow {\cal D}}_{x_i}{L}(x_i, x_{i+1},k)_{A_i}{\overleftarrow {\cal D}}_{x_{i+1}}
\; G(k)_{A_1\dots A_n} \ ,
\label{general_form}
\ee
where ${\cal D}$ are differential operators which are present if ${\cal O}_i$ contain derivatives and
\be
{L}(x_1, x_2,k)_{A_i}=\int d^dp_1 d^dp_2 \, e^{ip_1\cdot x_1 + ip_2\cdot x_2 }
\, {\widetilde L}(p_1, p_2,k)_{A_1} \ .
\ee
The presentation (\ref{general_form}) of correlation functions is quite general and does
not assume any specific structure for the operators ${\cal O}_i$ apart from their
two-field structure. In the following we will mainly restrict to operators carrying
nontrivial R-charge. 
By considering all possible choices of $L(x_1, x_2, k)$ we will identify the one that is dominant
-- {\it i.e.} it has the strongest singularity -- in the limit $|x_{12}|^2\rightarrow 0$. We will
moreover see that this $L(x_1, x_2, k)$ may be interpreted in terms of a null Wilson line with
fields attached to it.

While this appears to be a classical computation, it may be promoted to an all-loop one
in a regularization scheme in which the (partial) null-separation limit can be decoupled
from the integrals over the internal momenta $k$.
Such a scheme indeed exists: it suffices to take the limit $|x_{i, i+1}|^2\rightarrow 0$ such
that\footnote{In dimensional regularization one also needs to assume that
$\mu^2 |x_{12}|^{2}\rightarrow 0$ in the null separation limit, where $\mu$ is the
mass scale introduced by dimensional regularization.}
\be
|x_{i, i+1}|^{2} k^2 \rightarrow 0
\label{def_scheme}
\ee
for all possible internal momenta $k$. With this assumption, the null separation limit
can be taken at the level of the (regularized) integrand. In the following sections we will show
that, in this case, a sequence of scalar propagators with gluons attached to in though three-point
vertices reduces in the null separation limit to a null Wilson line with the same number of
attached gluons; this yields the desired results, to all orders in weak coupling perturbation
theory. As discussed in \cite{AEKMS}, in other schemes this is only a proportionality
relation.

As outlined here, the arguments in the following sections focus on only two operators at a 
time and justify the appearance of a Wilson line between their insertion points in the limit 
in which they are null-separated. There arguments will also support the relation between 
correlation functions in the limit in which $x_{i,i+1}^2=0$ but $x_{i-1,i}^2\ne 0$ for some 
subset of operators and the correlation function of appropriately capped open null zig-zag 
Wilson lines, as illustrated in eq.~(\ref{corropenWL}).

\section{2-field operators; no insertions \label{no_insertions}}

With the strategy outlined in the previous section, let us now proceed to prove the first equality
(\ref{allKS}) for charged 2-field BPS operators to all orders in perturbation theory and in the
presence of a dimensional regulator.
We will begin by showing that, in the limit of null separation,
the diagrams exhibiting a connected R-charge flow dominate over the diagrams with disconnected
flows and that the former reduce to the expectation value of a Wilson loop in the adjoint representation.
Moreover, it will turn out that the dominant diagrams will contain a continuous sequence of scalar
propagators.

\subsection{R-charge flow through scalar exchange \label{sgv}}

\begin{figure}[ht]
\centering
\includegraphics[height=19mm]{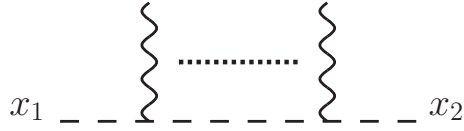}
\caption{Contributions of scalar propagators (represented by dashed lines) and
gluon-scalar vertices.  \label{manyssg}}
\end{figure}

Let us consider a sequence of $n$ one-gluon vertices connected by scalar propagators,  as shown in fig.~\ref{manyssg}. The momenta
going to the two endpoints are denoted $p_1$ and $p_2$ while the momenta carried by gluons
are denoted by $k_i$ and the momenta of the scalars between vertices are denoted
by $q_i$. Throughout we will not write explicitly the propagator of the gluons attached to
the vertices between the points $x_1$ and $x_2$.
The starting expression is thus:
\be
&&
L(x_1, x_2, k_1,\cdots, k_n)_{\mu_1\cdots\mu_n}=\int d^dp_1\int d^dp_2e^{ip_1\cdot x_1+ip_2\cdot x_2}\int d^dq_1\cdots\int d^dq_{n-1}\nonumber\\
&&\qquad\quad
\times\frac{(p_1-q_1)_{\mu_1}(-q_1-q_2)_{\mu_2}\cdots(-q_{n-2}-q_{n-1})_{\mu_{n-1}}(-q_{n-1}-p_2)_{\mu_{n}}}{(p_1^2+i0)(q_1^2+i0)\cdots(q_{n-1}^2+i0)(p_2^2+i0)}\label{many_gss_ini}\\
&&
\times\delta^{(d)}(k_1+q_1+p_1)\delta^{(d)}(k_2-q_1+q_2)\cdots\delta^{(d)}(k_{n-1}-q_{n-2}+q_{n-1})\delta^{(d)}(k_n-q_{n-1}+p_2)\nonumber
\ee

By performing all the integrals except the one over $p_1$, Schwinger-parametrizing
the propagators and representing the factors of $p_1$ in the numerator as derivatives
with respect to  $x_1$ it follows that
\be
&&L(x_1, x_2, k_1,\cdots, k_n)_{\mu_1\cdots\mu_n}
=\int d^dp_1
\frac{e^{ip_1\cdot (x_1-x_2)-i\sum_ik_i\cdot x_2}
\prod_{i=1}^n(2p_1+2\sum_{j=1}^{i-1}k_j+k_i)_{\mu_i}}{(p_1^2+i0)
\prod_{i=1}^n((p_1+\sum_{j=1}^ik_j)^2+i0)}\nonumber\\
&&\qquad\qquad
={}(-i)^{(n+1)}\prod_{i=1}^n
\left(-2i\frac{\partial}{\partial x_1^{\mu_i}}+2\sum_{j=1}^{i-1}k_{j\mu_i}+k_{i\mu_i}\right)
\left(\prod_{i=1}^{n+1}\int_0^\infty d\alpha_i\right)e^{-i\sum_ik_i\cdot x_2}\nonumber\\
&&\qquad\qquad
\qquad\qquad\qquad\qquad
\times\int d^dp_1e^{ip_1\cdot(x_1-x_2)+i\sum_{i=1}^{n+1}\alpha_i(p_1+\sum_{j=1}^{i-1}k_j)^2-0\sum\alpha_i} \ .
\ee
%
%
The change of variables  $\alpha_i=\zeta s_i$ with $s_i\in [0, 1)$, $\sum_{i=1}^{n+1} s_i=1$
and $\zeta\in [0,\infty)$ further simplifies this expression.
Moreover, writing $s_i=t_i-t_{i-1}$ for $i=1,\dots,n$ and with $t_0=0$, the unit sum constraint
on the $s$ variables becomes just $s_{n+1}=1-t_n$. Together with the evaluation of the $p_1$ integral, these transformations lead to:
\be
\label{many_gss}
&&L(x_1, x_2, k_1,\cdots, k_n)_{\mu_1\cdots\mu_n}
=(-i)^{(n+1)}\prod_{i=1}^n
\left(-2i\frac{\partial}{\partial x_1^{\mu_i}}+2\sum_{j=1}^{i-1}k_{j\mu_i}+k_{i\mu_i}\right)
\\
&&\quad
\times\left(\prod_{i=1}^{n}\int_{t_{i-1}}^{1} dt_i\right)e^{-i\sum_ik_i\cdot (x_2t_i-x_1(1-t_i))}
\int_0^\infty d\zeta\zeta^n\frac{\pi^{2-\epsilon}}{(-i\zeta)^{2-\epsilon}}e^{i\zeta
\tilde{f}(t_i,k_i)-i\frac{(x_1-x_2)^2}{4\zeta}-0\zeta-0/\zeta} \ ,~
\nonumber
\ee
with some functions $\tilde{f}(t_i,k_i)$ whose expressions are not important.

The integral over $\zeta$ is of the general type
\be
\label{allints}
I_m(z, f)=\int d\zeta\; \zeta^{m-2+\epsilon}e^{i\zeta f-i\frac{z^2}{\zeta}-0\zeta-0/\zeta}
\ee
with some choice of $m$ and with some function $f$ depending on the momenta $k_i$
and the affine parameters $t_i$; such integrals will also appear in later sections in
diagrams involving other types of fields.
The rescaling $\zeta\rightarrow \zeta z^2$ together with the
null-separation limit implies that
\be
\lim_{z^2\to 0} I_m(z, f)
=
(-1)^{\frac{1-m-\epsilon}{2}}\Gamma(1-m-\epsilon)z^{2(m+\epsilon-1)}
\label{generic_int}
\ee
for $m\le 0$. For $m\ge 1$ the integral does not vanish, but reduces to an $f$-dependent
(and $z$-independent) constant.

The leading term in the light-like limit arises from the highest power of $\zeta^{-1}$
brought down by the differentiation with respect to $x_1$. It is easy to see that it is
$\zeta^{-n}$; this factor cancels the $\zeta^n$ measure factor arising form the change of
variables and leaves behind an $n$-independent $\zeta$ factor which generates, upon
integration, a factor of the position space scalar
propagator $\Delta_{4-2\epsilon}(x_{1}-x_2)$.\footnote{As usual, the position
space scalar propagator is just
\be
\nonumber
\Delta_d(x_1-x_2)&=&\int d^dp_1d^dp_2\frac{e^{ip_1\cdot x_1+ip_2\cdot x_2}}{p_1^2+i0}
\delta^{(d)}(p_1+p_2) \\
&=&{}-i\int_0^\infty d\zeta\int d^dp_1e^{ip_1\cdot(x_1-x_2)+i\zeta p_1^2-\zeta0}
={}-i\int_0^\infty d\zeta\frac{\pi^{2-\epsilon}}{(-i\zeta)^{2-\epsilon}}e^{-i\frac{(x_1-x_2)^2}{4\zeta}-0/\zeta} \ ,
\nonumber
\ee
where $\zeta$ is introduced here as a Schwinger parameter.
}
The remaining integral represents  $n$ correctly ordered Wilson line vertices together with the
exponential factors needed to Fourier-transform the gluon propagator:
\be
&&L(x_1, x_2, k_1,\cdots, k_n)_{\mu_1\cdots\mu_n}\cr
&&\qquad\qquad
=\Delta_{4-2\epsilon}(x_{1}-x_2)\,
\prod_{i=1}^n\;i(x_1-x_2)_{\mu_i}\;
\int_{t_{i-1}}^{1} dt_i\;e^{-i\sum_ik_i\cdot (x_2t_i-x_1(1-t_i))}~,~~~~
\label{many_gss_final}
\ee
where, as mentioned before, $t_0=0$.
We have therefore recovered, in the scheme used here, the results of the eikonal
approximation: a continuous sequence of scalar propagators connecting two operators
that become light-like separated is equivalent to a light-like Wilson line between the
operators' positions.~\footnote{This is consistent  with the scalar field self-energy
vanishing in this scheme. Indeed, contracting two of the free indices  one finds an
additional $x_{12}^2$ factor which makes the right-hand side of  eq.~(\ref{many_gss_final})
vanish in the null separation limit.}

The derivation described above points to a simple rule for identifying the diagrams that
survive in the light-like limit: once all propagators are written with standard $(p^2+i0)$
denominators, one simply counts the number of momentum factors of lines connecting
the two insertion points. If this number is equal to the number of propagators minus one,
then the corresponding diagram yields a sufficiently singular contribution. Otherwise it
drops out in the null separation limit.\footnote{Indeed, in equation (\ref{many_gss}) the
measure factor $\zeta^n$ counts the number of propagators minus one and the derivative
factors count the number of momenta along the path connecting the two insertion points.
Each derivative generates a $\zeta^{-1}$ factor; to have a sufficiently singular contribution
it is necessary, according to (\ref{generic_int}), that these two factors cancel out or yield
some negative power of $\zeta$.  }

For this reason the four-point gluon-scalar interaction cannot contribute to the leading
term in the null-separation limit: since such a vertex has no derivatives, replacing a three-point
vertex in eq.(\ref{many_gss_ini}) with one such vertex would lead to an eq.~(\ref{many_gss})
with one fewer derivative factor and hence to an extra factor of $\zeta$ in the integral.
It will therefore not be sufficiently singular to contribute in
the limit in which R-charge flows between light-like separated points.

\subsection{Other interactions leading to R-charge flow}

\begin{figure}[ht]
\begin{center}
\includegraphics[height=18mm]{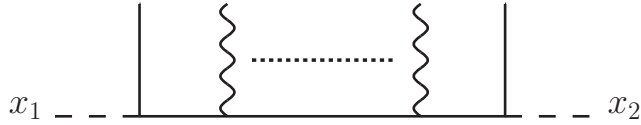}
\caption{An example of diagram exhibiting R-charge flow which is subleading in the null-separation limit. Dashed lines represent scalar fields while solid lines represent fermions.  \label{many_and_sff}}
\end{center}
\end{figure}

Apart from a continuous sequence of scalar propagators,
R-charge flow  between two operators may be realized by a sequence
of scalars and fermions. We illustrate this possibility in  fig.~\ref{many_and_sff} with
two scalars and any number of fermions; the sequence of fermion propagators may also be
interrupted by further scalar lines, in which case more of the open-ended lines become scalars.

The relation between the singularity in the null separation limit and the structure of the
momentum dependence of the integrand of the Fourier transform of such a sequence  suggests that,
for a fixed number of propagators, the most singular terms arise from the terms with the highest
number of momentum factors in the numerator.
This illustrates the importance of the spin of the exchanged particle in the null-separation limit.
Since no derivatives appear in the interaction of fermions and scalars and of fermions and gluons,
the only source of numerator momenta in fig.~\ref{many_and_sff} is the numerator of fermion
propagators. Consequently, the diagrams with more than the two scalar propagators will have a
softer singularity than the diagram shown in this figure.
Let us then discuss the situation of the highest number of fermion propagators --- $(n-2)$ for a
total of $n$ lines.  The relevant  Fourier-transform is
\be
&&L(x_1,x_2, k_1,\cdots, k_{n})_{\mu_1\cdots\mu_{n-2}}\cr
&=&\int d^d p_1 \int d^d p_2 e^{ip_1\cdot x_1+ip_2\cdot x_2}
\int d^d q_1\dots \int d^d q_{n-1}
\frac{q\llap/{}_{n-1}\gamma_{\mu_{n-2}}\dots q\llap/{}_2\gamma_{\mu_1} q\llap/{}_1}
{(p_1^2+i0)(q_1^2+i0)\dots (q_{n-1}^2+i0)(p_2^2+i0)}\nonumber\\
&&
\times\delta^{(d)}(k_1+q_1+p_1)\delta^{(d)}(k_2-q_1+q_2)\cdots
\delta^{(d)}(k_{n-1}-q_{n-2}+q_{n-1})\delta^{(d)}(k_n-q_{n-1}+p_2)~.~~
\ee
Carrying out the $q$ integrals, Schwinger-parametrizing the resulting propagators and changing
variables as discussed in the previous section leads to an expression of the type
\be
&&L(x_1,x_2, k_1,\cdots, k_n)_{\mu_1\cdots\mu_{n-2}}\cr
&\sim&\left[\prod_{j=1}^{n-1}\frac{\partial}{\partial x_1^{\sigma_j}}+{\rm fewer ~derivatives}\right]
\left(\prod_{i=1}^{n}\int_{t_{i-1}}^{1} dt_i\right)e^{-i\sum_ik_i\cdot (x_2t_i-x_1(1-t_i))}\nonumber\\
&&\qquad
\times T^{\sigma_{n-1}\mu_{n-2}\dots\mu_1\sigma_1}\int_0^\infty d\zeta\zeta^n\frac{\pi^{2-\epsilon}}{(-i\zeta)^{2-\epsilon}}e^{i\zeta
\tilde{f}(t_i,k_i)-i\frac{(x_1-x_2)^2}{4\zeta}-0\zeta-0/\zeta}~~,
\ee
where $T$ is a product of Dirac matrices.
It is easy to see that, among the integrals defined in (\ref{allints}), only $I_m$ with $m\ge 1$
can appear from the action of derivatives.
As expected, it follows from (\ref{generic_int}) that all such diagrams are not sufficiently
singular in the limit of null separation and thus can be ignored.

It is clear that a similar argument goes through if some of the fermion lines are replaced with scalar lines;
such contributions can also be ignored.

\subsection{No R-charge flow between two insertion points \label{noR}}

\begin{figure}[ht]
\begin{center}
\includegraphics[height=19mm]{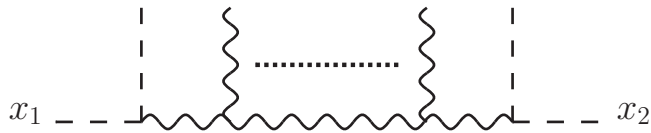}
\caption{Example of contributions without R-charge flow between two operators.\label{noflow}}
\end{center}
\end{figure}

The typical diagram that does not exhibit R-charge flow between two charged operators
involves at least one gluon propagating along any possible path between them -- see fig.~\ref{noflow}.
The sequence of gluon propagators may be interrupted by scalar or fermion lines.
Since the fermion-gluon interaction has no derivatives, the discussion in the previous section implies
that such diagrams are not sufficiently singular in the null-separation limit and we can ignore them.
We will therefore focus on gluons and scalars and, for the same reason as above, also ignore
the four-point interactions.

The momentum dependence of the gluon-scalar three-point vertices guarantees that
Feynman diagrams of the type shown in fig.~\ref{noflow} have the correct momentum
dependence to potentially yield leading order contributions in the limit of null separation
of insertion points; this would be the case if, upon Fourier transform, each vertex contributes
a factor of $(x_1-x_2)^\mu$. Such a  relation also appeared in \S~\ref{sgv}.
A closer inspection reveals however that these potentially dangerous diagrams do not contribute.
To see this let us assume that there are $n$ vertices and $(n+1)$ propagators between insertion points.
Of the external lines not attached to the insertion points, at least two are scalars, leaving only
at most $(n-2)$ outgoing gluon lines and consequently at most $(n-2)$ free Lorentz indices.
Since there are $n$ factors of $(x_1-x_2)^\mu$, one for each vertex, it immediately follows that at
least two such factors are necessarily contracted (otherwise there would be more than $n-2$ free indices)
and at least one more $(x_1-x_2)^2$ numerator factor is generated.\footnote{The discussion here assumes
that the gluon propagators are in Feynman gauge. The same result may be found in other gauges;
even though one gets more possibilities for contracting the different vectors as it also brings about
an extra factor of $q^{-2}$, {\it i.e.} additional propagator-like factors.}

Since the Feynman diagrams contributing to fig.~\ref{noflow} depend on a number of additional
vectors -- {\it e.g.} the momenta of external particles -- one may wonder whether two of the
indices arising from the $n$ vertices  may be contracted with such vectors rather than produce an
$(x_1-x_2)^2$. Such contributions can be ruled out by noticing, on the one hand dimensional analysis
requires that the number of numerator factors be fixed\footnote{That is, one may at most replace a
factor of $x_1-x_2$ by a factor of $k$.} and on the other factors of external momenta lower the singularity
through the absence of $\zeta^{-1}$ factors.

It therefore follows that, regardless of the mechanism that is responsible for the reduction of
the number of free vector indices -- either contraction with some $k_i$ or the appearance of an explicit
$(x_1-x_2)^2$ factor -- the null separation limit $(x_1-x_2)^2\rightarrow 0$ contains no singularities.
This completes the proof that, if a connected R-charge flow can exist between scalar operators,
then only Feynman diagrams leading to this  flow contribute in the null separation limit.

We have therefore shown that, in the presence of a dimensional regulator and in the scheme
described in \S~\ref{generalities}, the correlation
function of length-2 scalar operators in the null separation limit is proportional to the
expectation value of a null polygonal Wilson loop in the adjoint representation with cusps at
the positions of the original operators:
\be
\langle {\cal O}(x_1)\cdots {\cal O}(x_n)\rangle = \langle W_n\rangle_{\text{adj}}
\;
\prod_{i=1}^n\Delta(x_{i}-x_{i+1})~.
\ee
In general, as explained in \cite{AEKMS, EKS}, a scheme-dependent coefficient function will
appear on the right-hand side of this equation and will capture the differences between different
possible light-like limits and their inter-relation with the regulator.
The arguments presented here hold for finite values of the rank of the gauge group and do 
not rely on the existence of supersymmetry. Our proof extends the arguments of 
\cite{ABMSproof} to the regularized theory. We will comment in the Appendix on
correlation functions of operators built from other fields.

\section{2-field operators and additional operator insertions \label{w_insertion}}

Let us now proceed to discuss more general limits of correlation functions, in which 
only a subset of operators participate in the null separation limit.

\subsection{A single additional operator}

Let us first consider the correlation functions
of $n+1$ operators, $n$ of which participate in the null separation limit. The additional
operator ${\cal O}(a)$ is located at a generic position, $(x_i-a)^2 \ne 0$ for all $i=1,\cdots, n$.
 As discussed in the Introduction, such an
operator should have vanishing R-charge, otherwise a nonvanishing correlation function
$\langle {\cal O}(x_1)\cdots {\cal O}(x_n)\rangle $ implies that
$\langle {\cal O}(x_1)\cdots {\cal O}(x_n){\cal O}(a)\rangle=0 $.

\begin{figure}[ht]
\begin{center}
\includegraphics[height=25mm]{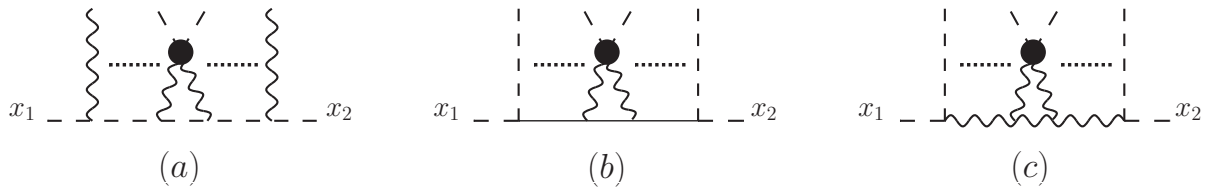}
\caption{Example of diagrams in which the additional operator (denoted by a heavy dot) acts
as a Lagrangian interaction vertex.
\label{extra1}}
\end{center}
\end{figure}

The arguments in the previous section apply equally well if the additional operator
does not affect the R-charge flow between two light-like separated operators. Indeed, if
${\cal O}(a)$ is connected to the scalar and/or fermion lines as in  fig.~\ref{extra1}(a) and
\ref{extra1}(b), then it acts similarly to any other multi-point interaction arising from the
Lagrangian: the continuous
sequence of scalar propagators will continue to dominate  in the presence of an additional operator
insertion. A similar conclusion -- that the presence of the additional operator does not affect the
conclusion of the previous section -- can be reached if the operators is inserted in the diagrams in
fig.~(\ref{noflow}) while being attached to some of the fields denoted by open lines -- see fig.~\ref{extra1}(c). Dimensional arguments also imply that diagrams that are subleading in the absence
of the additional operator cannot acquire a stronger singularity.

\begin{figure}[ht]
\begin{center}
\includegraphics[height=21mm]{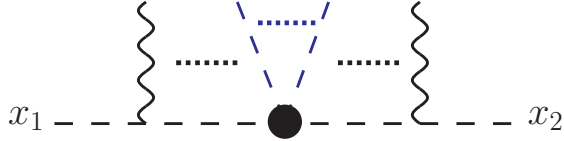}
\caption{An operator insertion (denoted by a heavy dot) along an R-charge flow. Since the insertion point is not
integrated over, the limit $(x_1-x_2)^2 \rightarrow 0$ is non-singular. \label{extra2}}
\end{center}
\end{figure}

The other possible way of inserting the additional operator is by  interrupting the sequence of
propagators explicitly shown in figs.~\ref{manyssg}, \ref{many_and_sff} and \ref{noflow}, as illustrated in fig.~\ref{extra2} for the Feynman diagrams dominant in the null separation limit.
%
One may intuitively expect that, since the insertion point $a$ is not null-separated from
$x_1$ and $x_2$, the limit $(x_1-x_2)^2 \rightarrow 0$ is non-singular\footnote{This
would no longer be the case if $a$ were integrated over.}.
A short calculation shows that this is indeed the case; we shall illustrate it with the example
of an operator ${\cal O}(a)$ with arbitrary number of derivatives inserted in a continuous sequence of
scalar propagators.

Assuming that the operator is inserted at position $a$, that the fields that are Wick-contracted
with the scalars in the original sequence of propagators carry $n_1$ and $n_2$ derivatives and
using $L(x_1,x_2, k)_{\mu_1\cdots\mu_n}$ in eq~(\ref{many_gss}) it is easy to see that
fig.~\ref{extra2} will contribute the following to the correlation function of the $n+1$ operators:
\be
&&L(x_1, a, x_2, k_1,\cdots, k_n, O')_{\mu_1\cdots\mu_{n}}\cr
&=&(\partial_a)^{n_1}
L(x_1,a, k_1,\cdots, k_r)_{\mu_1\cdots\mu_{r}}
\;{{\widetilde O}'(a)}\;(\partial_a)^{n_2}
L(a,x_2, k_{r+1},\cdots, k_n)_{\mu_{r+1}\cdots\mu_{n}} \ ,
\ee
where $(\partial_a)^{n_1}$ and $(\partial_a)^{n_1}$ stand for products of derivatives with
potentially suppressed free indices and ${{\widetilde O}'} $ contains the fields of
${\cal O}(a)$ that are not contracted with the scalar line.
Neither one of the two $L$ factors is singular as $(x_1-x_2)^2\rightarrow 0$. It therefore follows
that these terms, as well as the others obtained by replacing some or all of the scalars with gluons and/or
fermions give subleading contributions compared to the Feynman diagrams in which $O$ is not
interrupting the sequence of scalar propagators connecting two light-like separated points.

Thus, in the limit in which $x_i$ with $i=1,\dots,n$ are sequentially null separated,
the correlator $\langle O(x_1)\cdots O(x_n)O(a)\rangle$ becomes proportional to the
correlation function of the null Wilson loop in the adjoint representation
with corners at positions $x_i$ and the additional operator ${\cal O}(a)$.

\subsection{Several additional operators}

It is not difficult to generalize the arguments in the previous section to the insertion of several
operators which are placed at generic positions relative to the points $x_i$ with $1\dots n$.
In such case the only constraint stemming from R-charge conservation is that the
total charge of the insertions vanishes: $\sum_{j=1}^M Q_R({\cal O}(a_m))=0$.

If none of the operators interrupts the R-charge flow between two null-separated operators they
act, as in the case of a single operator insertion, as a Lagrangian interaction vertex. If a sequence
of scalar propagators is interrupted by an additional operator then it becomes non-singular in
the limit in which its beginning and end points are null separated. It therefore
follows that
\be
\label{partialnull}
\lim_{x_{ii+1}^2\to 0}
\frac{\langle {\cal O}(x_1)\cdots {\cal O}(x_n){\cal O}(a_1)\cdots{\cal O}(a_m)\rangle^{\text{conn}}}
{\langle {\cal O}(x_1)\cdots {\cal O}(x_n)\rangle}
=
\frac{\langle W_n\,{\cal O}(a_1)\cdots{\cal O}(a_m)\rangle^{\text{conn}}}
{\langle W_n\rangle} ~~.
\ee
As before, the Wilson loop is in the adjoint representation and the upper index denotes the
fact that on both sides of this equation one should restrict to the connected part of the correlation
functions. Relaxing this constraint will replace the right-hand side numerator with a sum over
connected components, each of which being the product of the correlation function of $W_n$
and some subset of
${\cal O}(a_1)\dots{\cal O}(a_m)$ and the correlation function of the remaining operators.

For sufficiently many additional operator insertions one may take a further
limit, in which $(a_j-a_{j+1})^2\rightarrow 0$ but at generic positions compared
to $x_{i}$. Dividing by $\langle {\cal O}(a_1)\cdots{\cal O}(a_m)\rangle$ it then
follows that
\be
\lim_{\stackrel{\scriptstyle{x_{ii+1}^2\to 0}}{\scriptstyle{a_{jj+1}^2\to 0}}}
\frac{\langle {\cal O}(x_1)\cdots {\cal O}(x_n){\cal O}(a_1)\cdots{\cal O}(a_m)\rangle}
{\langle {\cal O}(x_1)\cdots {\cal O}(x_n)\rangle\langle {\cal O}(a_1)\cdots{\cal O}(a_m)\rangle}
=1+
\frac{\langle W_n\,W_m\rangle_{\text{adj}}^{\text{conn}}}{\langle W_n\rangle\langle W_m\rangle_{\text{adj}}} \ .
\label{relWW}
\ee
As before, the upper index on the right-hand side denotes a restriction to the connected part of that
correlator; the constant term is the contribution of one of the disconnected components of
$\langle {\cal O}(x_1)\cdots {\cal O}(x_n){\cal O}(a_1)\cdots{\cal O}(a_m)\rangle$. The other disconnected components have a different singularity structure and their contribution to the ratio
(\ref{relWW}) vanishes in the null separation limit.

One may consider the factorization of the correlation function of Wilson loops, as outlined in
eq.~(\ref{factorization}). From the perspective of eq.~(\ref{relWW}), it corresponds to
a particular
factorization of the correlation function of $n+m$ operators into the sum of products of $n+1$ and
$m+1$-point correlation functions.

\subsection{Large $N$ limit}

\begin{figure}[ht]
\begin{center}
\includegraphics[height=30mm]{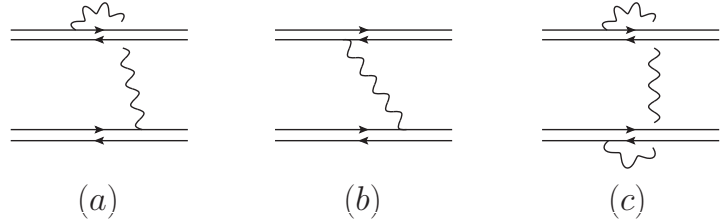}
\caption{Leading and subleading diagrams in the large-$N$ limit. The double-lines denote
Wilson lines and the wavy line represent gluons.   \label{largeN}}
\end{center}
\end{figure}

The arguments described above are independent of the rank of the $SU(N)$ gauge group. To
make contact with a strong coupling analysis it is necessary to consider the large-$N$ limit. In the
absence of operator insertions this limit is standard
\be
\lim_{N\to \infty} \langle W_n\rangle_{\text{adj}} = \langle W_n\rangle_{\text{fund}}^2 \ .
\label{fact_adj}
\ee
Diagrammatically, this relation arises from the fact that the Feynman diagrams of the type shown in fig.~(\ref{largeN})(a) are subleading in the large $N$ limit compared to diagrams in
figs.~(\ref{largeN})(b). The graph in fig.~\ref{largeN}(c) may or may not contribute depending on the rest of the diagram.
That is, the only Feynman diagrams that survive in this limit are those in which the inner and
outer index lines of the Wilson loop are connected to themselves by some webs of vertices and
propagators from the Lagrangian. Moreover, away from the Wilson loop, the Feynman diagram is planar.

It is not difficult to generalize this picture to the correlation function of a Wilson loop and an operator,
$\langle W_n {\cal O}(a)\rangle_{\text{adj}}$. The Feynman diagrams that survive the large-$N$ limit
are those in which the operator ${\cal O}(a)$ is connected only to one index line of the Wilson loop. Thus,
\be
\lim_{N\to \infty} \langle W_n\,{\cal O}(a)\rangle_{\text{adj}} =
2\langle W_n\rangle_{\text{fund}}\langle W_n{\cal O}(a)\rangle_{\text{fund}} \ .
\ee
The factor of $2$ arises because the operator may be connected either to the inner of the outer index
line ({\it e.g.} it may be inserted on the gluon line in figs.~(\ref{largeN})(b)~and~(c)). This factor is in
agreement with the discussion in \cite{ABT} on the insertion of the (integrated) dilaton vertex operator.

Following a similar reasoning it is easy to take the planar limit for the correlator of a Wilson loop and several operators. The  Feynman graphs can be organized into sets of diagrams in which some operators are connected to one Wilson loop index line and the other operators are connected to the other Wilson loop index in a planar way, {\it i.e.}
\be
\label{opsNinfty}
&&
\lim_{N\to \infty} \langle W_n\,{\cal O}(a_1)\cdots{\cal O}(a_m)\rangle_{\text{adj}}\\
&=&
\sum_{k=0}^m\sum_{\{j_1,\dots, j_k\}\subset \{1,\dots, m\}}
\langle W_n\,{\cal O}(a_{j_1})\cdots{\cal O}(a_{j_k})\rangle_{\text{fund}}
\langle W_n\,{\cal O}(a_{j_{k+1}})\cdots{\cal O}(a_{j_{m}})\rangle_{\text{fund}}
\nonumber\ ,
\ee
where the second sum runs over all subsets  of $\{1,\dots,m\}$ with $k$ elements.
If the total R-charge of the operators appearing
in an expectation value is nonzero, $\sum_{l=1}^kR({\cal O}(x_{j_l}))\ne 0$, the corresponding term vanishes identically.
For example, if $m=2$ the R-charges of the two operators must be equal in absolute value;
if their common value is nonvanishing, then all correlators involving one Wilson loop and
one operator vanish identically and, in the sum above, only the terms with $k=0$ and
$k=m=2$ survive.

Last, the factorization of the connected part of the correlation function of two Wilson loops follows a
similar pattern:
\be
\lim_{N\to\infty}\langle W_nW_m\rangle^{\text{conn}}_{\text{adj}} = 2\langle W_n\rangle_{\text{fund}}
\langle W_m\rangle_{\text{fund}}\langle W_nW_m\rangle^{\text{conn}}_{\text{fund}} \ .
\label{WWlim}
\ee
This term arises form the $k=0$ and $k=m$ terms in eq.~(\ref{opsNinfty}). The other terms disappear in
the null limit because there is no R-charge flow between operators (in fact there are no fields that
propagate between the operators in one set and the operators in the other) and thus are subleading compared
to eq.~(\ref{WWlim}).

\section{Operators with higher spin \label{large_charge}}

As reviewed in \S~\ref{intro}, it was suggested in \cite{AEKMS} that the relation
between correlation functions and
Wilson loops is not restricted to the operators in the stress tensor multiplet.
If such a generalization is indeed correct, it is tempting to expect that it would hold
in the presence of additional insertions as well.
Such a generalization, both in the absence and in the presence of additional operator insertions,
is potentially very interesting. A relation between correlation functions and Wilson loops suggests
that, at least in the null separation limit and at string coupling, correlation functions should
have a semiclassical description.
While such an approach is difficult to justify for operators of small dimension (even in
the null separation limit), a justification is readily available
\cite{3PFsemiclassics}
for operators whose dimension scales like $\Delta \propto \sqrt{\lambda}$.

\subsection{No additional insertions}

The discussion in the previous sections can be extended without much difficulty to the
case when the operators inserted at positions $x_1,\cdots, x_n$ are twist-2 operators
\be
{\cal O}^S=\sum_{n=0}^S \,c_n(\lambda)\Tr[D_+^nZD_+^{S-n}Z] \ ,
\label{opS}
\ee
where the coefficients $c_n(\lambda)$ are determined by requiring that these operators have definite
anomalous dimensions. At one loop, they are given \cite{Belitsky:2003ys} in terms of the Jacobi polynomials
\be
{\cal O}^S = i^s(n\cdot D_{1}+n\cdot D_{2})^SP_S^{(0,0)}
\left(\textstyle{\frac{n\cdot D_{2}-n\cdot D_{1}}{n\cdot D_{2}+n\cdot D_{1}}}\right)
\Tr[Z(\xi_2)Z(\xi_1)]\Big|_{\xi_1=\xi_2} \ ,
\ee
where $n\cdot D$ are covariant derivatives in the adjoint representation in the light-like direction specified by the vector $n$.
Our regularization scheme implies however that the following arguments hold even if
${\cal O}^S$ does not have definite anomalous dimension.

As we saw in the previous sections, in the null separation limit, the Feynman diagrams that
dominate the correlation function of 2-field scalar operators generate R-charge flow between
insertion points and contribute the scalar propagator  $\Delta(x_1-x_2)$ multiplied by a Wilson line
stretched between $x_1$ and $x_2$, see eq.~(\ref{many_gss}).
The  additional derivatives present in the operators~(\ref{opS}) act either on the scalar propagator
prefactors $\Delta(x_i-x_{i+1})$ or on the Wilson line factor in the usual way:
\be
\langle (n\cdot D)^pZ(x_1)(n\cdot D)^mZ(x_2)\rangle\mapsto (n\cdot D_{x_1})^p(n\cdot D_{x_2})^m
\langle Z(x_1)Z(x_2)\rangle \ .
\ee
The derivative action on $\Delta(x_i-x_{i+1})$ yields terms that are more singular than
$\Delta(x_i-x_{i+1})$. Since the action of derivatives on the Wilson line does not generate
any $1/(x_i-x_{i+1})^2$ singularities, it follows that the most singular term in the null separation
limit arises solely from all derivatives acting on the scalar propagator prefactors:
\be
\nonumber
\lim_{x_{i,i+1}\to 0}\langle{\cal O}^{S_n}(x_n)\cdots {\cal O}^{S_1}(x_1)\rangle
&=&\langle W_n\rangle \sum_{m_1=0}^{S_1}\cdots \sum_{m_n=0}^{S_n}c_{m_1}\cdots c_{m_n}
\prod_{i=1}^n{\overleftarrow{n\cdot \partial}}_{x_{i}}{\overrightarrow{n\cdot \partial}}_{x_{i}}
\Delta(x_i-x_{i+1})
\\
&& \hspace{-20truemm}
+(\text{contributions of the gauge fields in covariant derivatives})
\label{corr_tw2}
\ee
where $x_{n+1}\equiv x_1$ and the left derivative with respect to $x_1^+$ in
the first term acts on the $\Delta(x_n-x_1)$ factor.

\begin{figure}[ht]
\begin{center}
\includegraphics[height=25mm]{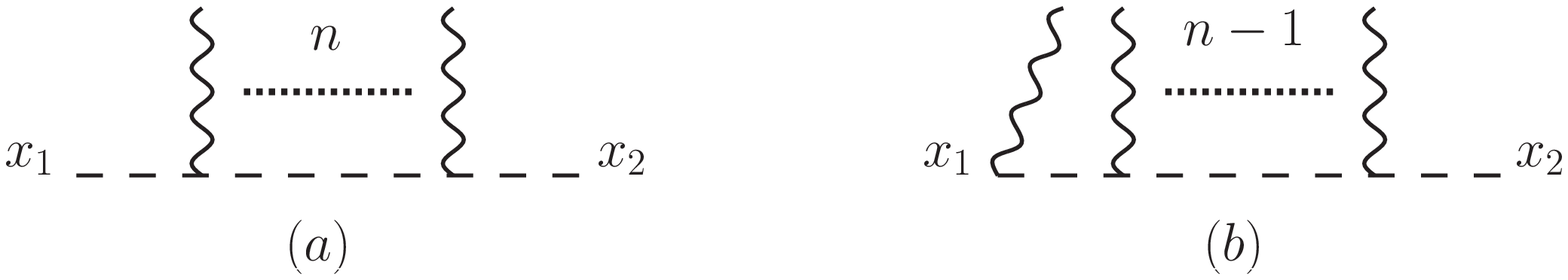}
\caption{Contribution of the gauge field inside the covariant derivative $(b)$ compared to the
contribution of a scalar line with the same number of external gluons $(a)$.\label{corner}}
\end{center}
\end{figure}

The contribution of the gauge fields present in the covariant derivatives can be shown to also be
subleading. To understand this it is useful to interpret the $[A,Z]$ term in the covariant derivative
as a standard $AZ{\bar Z}$ vertex from which one strips off the derivative. Thus, for the purpose of
understanding the $x_{12}^2$ dependence of the gauge field contribution, we may simply analyze
the expression in eq.~(\ref{many_gss_ini}) without the $1/p_1^2$ propagator and without the momentum
contribution of the first gluon-scalar vertex -- see fig.~\ref{corner}(b).
It is useful to recall that the final $(x_{1}-x_2)$ dependence is governed by the power of $\zeta$ in
equation~(\ref{many_gss}); there, $\zeta^n$ in the measure arose from the propagators and each
derivative generates a factor of $\zeta^{-1}$. Thus, the two diagrams in fig.~\ref{corner} will have
identical singularities in the $(x_{1}-x_2)^2\to 0$ limit.
This singularity however is softer than that of the diagram in which the derivative acts on the scalar field; consequently, we may ignore the contribution of the gauge fields in the covariant derivatives. Thus, eq.~(\ref{corr_tw2}) reduces to
\be
\lim_{x_{i,i+1}\to 0}\langle{\cal O}^{S_1}(x_1)\cdots {\cal O}^{S_n}(x_n)\rangle
=\langle W_n\rangle \sum_{m_1=0}^{S_1}\cdots \sum_{m_n=0}^{S_n}c_{m_1}\cdots c_{m_n}
\prod_{i=1}^n{\overleftarrow{n\cdot \partial}}_{x_{i}}{\overrightarrow{n\cdot \partial}}_{x_{i}}
\Delta(x_i-x_{i+1})
\ .
\ee

\subsection{Additional operator insertions}

It is straightforward to extend the arguments above, along the lines of \S~\ref{w_insertion}, to
the correlation function of $n+1$ operators, $n$ of which are taken to be null separated. We find
\be
&&\lim_{x_{i,i+1}\to 0}\langle{\cal O}^{S_1}(x_1)\cdots {\cal O}^{S_n}(x_n) {\cal O}^{S_a}(a)\rangle
=
\langle W_n\,{\cal O}^{S_a}(a)\rangle_{\text{adj}}\\
&&\qquad\qquad\qquad\qquad\qquad\qquad
\times\sum_{m_1=0}^{S_1}\cdots \sum_{m_n=0}^{S_n}c_{m_1}\cdots c_{m_n}
\prod_{i=1}^n
{\overleftarrow{n\cdot \partial}}_{x_{i}}
{\overrightarrow{n\cdot \partial}}_{x_{i}}\Delta(x_i-x_{i+1})
\nonumber
\ .
\ee
Normalizing this expression to the correlation function of the $n$ null-separated operators it follows
immediately that
\be
\lim_{x_{i,i+1}^2\to 0}\frac{\langle{\cal O}^{S_1}(x_1)\cdots
{\cal O}^{S_n}(x_n) {\cal O}^{S_a}(a)\rangle}
{\langle{\cal O}^{S_1}(x_1)\cdots {\cal O}^{S_n}(x_n)\rangle}=
\frac{\langle W_n\,{\cal O}^{S_a}(a)\rangle_{\text{adj}}}{\langle W_n\rangle_{\text{adj}}} \ .
\ee
For several operator insertions at generic positions one finds for the connected part of
the correlation function that
\be
&&\lim_{x_{i,i+1}\to 0}\frac{\langle{\cal O}^{S_1}(x_1)\cdots {\cal O}^{S_n}(x_n)
{\cal O}^{S_{a_1}}(a_1)\dots {\cal O}^{S_{a_m}}(a_m)\rangle^{\text{conn}}}
{\langle{\cal O}^{S_1}(x_1)\cdots {\cal O}^{S_n}(x_n) \rangle}
\\
&&\hspace{80truemm}
=
\frac{\langle W_n\,{\cal O}^{S_{a_1}}(a_1)\dots {\cal O}^{S_{a_m}}(a_m)\rangle_{\text{adj}}^{\text{conn}}}
{\langle W_n \rangle_{\text{adj}} }\ ,
\nonumber
\ee
in close analogy with the correlation function of two-field BPS operators.

\section{Remarks \label{comments}}

Let us comment here on correlation functions of higher-twist operators.
In \cite{EKS} it was suggested and demonstrated through two loops for four-point correlators,
that the relation between correlation functions and null polygonal Wilson loops is not restricted
to two-field BPS operators or, more generally, operators in the stress tensor multiplet.
To see this in the regularization scheme we employed here let us consider operators
connected by two or more non-overlapping sequences of propagators allowing for R-charge
flow. Following the discussion in \S~\ref{no_insertions}, in the limit in which the two operators
are null-separated, each such sequence is equivalent to a Wilson line multiplied by a position
space scalar. Moreover, we have also seen that if fields other than gluons are attached to the
Wilson line, the singularity of the scalar propagator is softened.

The same holds if two Wilson lines between the same two null-separated points are connected
by a gluon.  Indeed, since the interaction vertex between a Wilson line and a gluon is proportional
to $x_{12}^\mu$, a gluon of 
this type  necessarily yields a factor of $|x_{12}|^2$
which softens the singularity of the scalar propagator.\footnote{This is consistent with the two-point function of operators being tree-level exact in this scheme.}
We therefore conclude that, if two operators are connected by more than one R-charge carrying
sequence of propagators, the leading contribution to the null separation limit comes from
Feynman  graphs in which no field connects two such sequences.
This in turn implies that, in the planar limit,
\be
\lim_{x_{i,i+1}^2\to 0}
\frac{\langle {\cal O}^{J_1}(x_1)\dots {\cal O}^{J_n}(x_n)\rangle}
{\langle {\cal O}^{J_1}(x_1)\dots {\cal O}^{J_n}(x_n)\rangle_0}
=
\frac{\langle W_n\rangle^2_{\text{fund}}}{\langle W_n\rangle^2_{0,\text{fund}}}
\label{allKSJ}
\ee
where ${\cal O}^{J_i}(x_i)$ stands for a scalar operator with $J_i$ fields. In general,
an additional scheme-dependent factor may appear on the right-hand side above. Unlike
the two-field scalar operators however, the argument above breaks down for a finite rank
gauge group.

This argument goes through unmodified if additional operators, placed at arbitrary positions,
are present in the correlation function. Indeed, as discussed in previous sections, if any of the
additional operators affect the R-charge flow between the null-separated ones, the resulting
Feynman diagram will have a subleading singularity. In the diagrams in which the additional
operators act as regular interaction vertices, the null separation limit proceeds as if they were
absent, implying that
\be
&&\lim_{\stackrel{\scriptstyle{x_{ii+1}^2\to 0}}{N\rightarrow\infty}}
\frac{\langle {\cal O}^{J_1}(x_1)\cdots {\cal O}^{J_n}(x_n)
{\cal O}(a_1)\cdots{\cal O}(a_m)\rangle^{\text{conn}}}
{\langle {\cal O}^{J_1}(x_1)\cdots {\cal O}^{J_n}(x_n)\rangle}\\
&&=
\frac{1}{\langle W_n\rangle_{\text{fund}}^2}
\sum_{k=0}^m\sum_{\{j_1,\dots, j_k\}\subset \{1,\dots, m\}}
\langle W_n\,{\cal O}(a_{j_1})\cdots{\cal O}(a_{j_k})\rangle_{\text{fund}}
\langle W_n\,{\cal O}(a_{j_{k+1}})\cdots{\cal O}(a_{j_{m}})\rangle_{\text{fund}}
\ .
\nonumber
\ee
Similarly to the partial null-separation limit of correlation functions of two-field
operators (\ref{partialnull}),
we expect that the scheme-dependent coefficient functions arising both in the numerator
and denominator of the left-hand side of this equation cancel each other out.

While the weak coupling arguments for the relation between correlation functions
and null polygonal Wilson loops are relatively straightforward, a detailed derivation of this relation remains mysterious from a strong coupling perspective. Moreover, as already mentioned in \cite{ABT} for two additional operators, if the charges and dimensions of the local operators are not large enough to invalidate the semiclassical expansion, the correlation function between a Wilson loop and some number of local operators -- such as the one on the right-hand side of eq.~(\ref{WW}) -- factorizes as
\be
\langle W_n{\cal O}_1(a_1)\dots {\cal O}_m(a_m)\rangle_{\text{fund}}
\rightarrow \prod_{j=1}^m\langle W_n{\cal O}_j(a_j)\rangle_{\text{fund}} \ .
\ee
If the operators ${\cal O}_j$ are the gauge theory dual of the  integrated dilaton vertex operators
({\it i.e.} they are the ${\cal N}=4$ action), a relation of this type can be understood at weak coupling \cite{BT3, RTcor} by simply representing \cite{Costaetal} ${\cal O}_j$ as $\lambda d/d\lambda$.
One may also expect that a similar relation may hold if the integrals over the positions of the
operators are omitted. It is however not immediately clear how to understand such a relation if
the operators ${\cal O}_j$ are other BPS operators;  it would be interesting to explore the origin
and  limitations of such a factorization at weak coupling.

Since the arguments discussed in \S~\ref{no_insertions} build the closed null polygonal 
Wilson loops one side at a time, they can be used to also prove the relation (\ref{corropenWL})
and its generalization to higher-point correlation functions and correlation function
of open null zig-zag Wilson lines with arbitrary number of cusps. Such relations may provide 
a semiclassical approach complementary to that of \cite{Roiban:2009aa,Tirziu:2008fk} to the 
calculation of anomalous dimensions of short operators, in particular of members of the Konishi 
multiplet.\footnote{To this end it is necessary to understand the string theory 
dual of open Wilson lines with fundamental fields at their ends.}

In this note we discussed in detail the relation between the null and partial null limits of
correlation functions and the expectation value of null polygonal Wilson loops as well
as the correlation functions of such Wilson loops and local gauge  invariant operators.
Twistor space techniques were used in \cite{bullimoreskinner} to construct a recursion
relation for the expectation value of null polygonal Wilson loops; this recursion relation
mirrors that of the unregularized integrand of scattering
amplitudes proposed in \cite{ArkaniHamed:2010kv}; see also ref.~\cite{Boels:2010nw}.
The two structurally different terms arise from the self-intersections of the Wilson loop due
to a BCFW-type deformation and from a quantum term in the loop equation, respectively
(see \cite{bullimoreskinner} for details). Through the construction in \cite{ABMSproof}
a similar recursion relation holds for correlation functions of gauge invariant operators in
the null-separation limit.
It would  be interesting to investigate the existence of similar recursion relations for partial
null-limits of correlation functions. It is not difficult to see that the BCFW-type
deformation will lead to additional terms arising form the intersection of the deformed Wilson
loop and the line(s) representing the insertion points of the additional operators. When such an
intersection occurs, one of the corners of the deformed Wilson loop becomes null-separated
from one of the additional operators. It may be possible that such contributions can also be
expressed in terms of null Wilson loops.

It may, in fact, be possible to construct BCFW-type
recursion relations for correlation functions away from null or partial null limits. Indeed, it
has been argued in \cite{Raju:2011mp} that strong coupling correlation functions in the
supergravity approximation obey such a relation. While the supergravity approximation
does not yield the desired null limit\footnote{{\it E.g.} the dependence of the 't~Hooft coupling
is not correctly captured. Other differences, related to the position dependence, are also present.},
the fact that  string amplitudes have a BCFW presentation
\cite{Boels:2008fc, Cheung:2010vn, Boels:2010bv} suggests that it may be possible to construct such a 
recursion relation also for strings in $AdS_5\times S^5$.

\bigskip

\section*{Acknowledgments}
We are grateful to  M.~Campiglia, G.~Korchemsky, D. Skinner and A. Tseytlin for discussions
and to G.~Korchemsky, P.H. Damgaard and A. Tseytlin for comments on a preliminary draft.
This work  is supported in part by the US Department of Energy under contract
DE-FG02-201390ER40577 (OJI)  and the A.P.~Sloan Foundation.


\bigskip
\bigskip

\appendix

\refstepcounter{section}
\def\theequation{A.\arabic{equation}}
\setcounter{equation}{0}

\section*{Appendix: Null separation limit for correlations of operators with
gauge fields and fermions \label{opotherfields}}

The arguments in \S~\ref{no_insertions} and \S~\ref{w_insertion} may be extended to correlation functions of 2-field operators containing fermions or gauge fields; one may interpret this as a step towards the proof of the supersymmetric generalization of the correlation function/Wilson loop relation. On dimensional grounds one expects that, since the engineering dimensions of such fields are larger than that of scalars\footnote{The gauge fields enter the operators as field strength factors.}, the correlators will be more singular in the null separation limit than the correlation function of scalar operators.

\subsection{2-field operators with fermions}

The position space fermion propagator is obtained by Fourier-transforming the momentum
space one in the standard way.
\be
\nonumber
S(x_1-x_2)&=&{}i{\partial\llap/}_1\int^\infty_0d\zeta\frac{\pi^{d/2}}{(-i\zeta)^{d/2}}
\exp\left[-i\frac{(x_1-x_2)^2}{4\zeta}-0/\zeta\right]\cr
&=&{}-(1-\epsilon)\pi^{2-\epsilon}2^{3-2\epsilon}\Gamma(1-\epsilon)
\frac{({x\llap/}_1-{x\llap/}_2)}{(x_1-x_2)^{2({2-\epsilon})}} \ .
\ee
Not unexpectedly, this propagator is more singular in the limit in which $x_1$ and $x_2$ are null-separated
than the scalar propagator.

Let us consider next a sequence of fermion propagators with gluons attached to it.
The important property of the fermion vertices is that they do not contain any derivatives.
However the momentum factor in the numerator of the fermion propagator will play a role
similar to the derivative in the scalar-gluon vertex. The contribution of such a sequence to the
expectation value of  operators is the factor
\be
&&L(x_1,x_2,k_1, \dots, k_n)_{\mu_1\dots\mu_n}\\
&=&\int d^dp_1\int d^dp_2e^{ip_1\cdot x_1+ip_2\cdot x_2}\int d^dq_1\cdots\int d^dq_{n-1}
\frac{ {p\llap/}_2\gamma^{ \mu_n} {q\llap/}_{n-1}
\cdots {q\llap/}_1\gamma^{ \mu_1} {{p\llap/}_1}}{(p_1^2+i0)(q_1^2+i0)\cdots(q_{n-1}^2+i0)(p_2^2+i0)}\nonumber\\
&&\times\delta^{(d)}(k_1+q_1+p_1)\delta^{(d)}(k_2-q_1+q_2)\cdots\delta^{(d)}(k_{n-1}-q_{n-2}+q_{n-1})\delta^{(d)}(k_n-q_{n-1}+p_2) \ .
\nonumber
\ee
Following the same steps as in the case of scalar operators we find that
\be
\label{procFermi}
&&L(x_1,x_2,k_1, \dots, k_n)_{\mu_1\dots\mu_n}\\
&=&{}(-i)^{(n+1)}(-1)^n\gamma^\nu\left(-i\frac{\partial}{\partial x_1^\nu}\right)
\prod_{i=1}^n\gamma^{ \mu_i}\gamma^{\mu_i}\left(-i\frac{\partial}
{\partial x^{\mu_i}_1}+\sum_{j=1}^{i}k_{j\mu_i}\right)\left(\prod_{i=1}^{n}
\int_{t_{i-1}}^{1} dt_i\right)\nonumber\\
&&\times e^{-i\sum_ik_i\cdot (x_2t_i-x_1(1-t_i))}
\int_0^\infty d\zeta\zeta^n\frac{\pi^{2-\epsilon}}{(-i\zeta)^{2-\epsilon}}e^{i\lambda\tilde{f}(t_i,k_i)-i\frac{(x_1-x_2)^2}{4\zeta}-0\zeta-0/\zeta}\nonumber \ .
\ee
The $\zeta$ integral is exactly of the same type as discussed before; therefore, the leading term in the
$(x_1-x_2)^2\to 0$ limit arises from the highest power of $\zeta^{-1}$ generated by the derivatives with
respect to $x$. Dirac matrix algebra may be used to simplify the numerator; discarding terms proportional to
$(x_1-x_2)^2$ we find that
\be
\label{finalFermi}
&&L(x_1,x_2,k_1, \dots, k_n)_{\mu_1\dots\mu_n}\\
&&\qquad\qquad\qquad
=i^{n+1}S(x_1-x_2)\left(\prod_{i=1}^n(x_1-x_2)^{\mu_i}\right)
\left(\prod_{i=1}^{n}\int_{t_{i-1}}^{1} dt_i\right)e^{-i\sum_ik_i\cdot (x_2t_i-x_1(1-t_i))} \ ,
\nonumber
\ee
{\it i.e.} a position space fermion propagator multiplied by a Wilson line with $n$ gluons attached to it.
This is the same structure as in the case of a sequence of scalar propagators.


There are two classes of diagrams which could potentially yield the same singularity as the
sequence of fermion propagators. One of them involves the fermion-scalar vertex. Having such a
vertex does not lead to a smaller number of numerator momenta. However, since this vertex
does not bring a vector index being proportional to only $\gamma^5$, its contribution at the
level of eq.~(\ref{procFermi}) will contain a factor of
\be
({x\llap/}_1-{x\llap/}_2)\gamma_5({x\llap/}_1-{x\llap/}_2)=-(x_1-x_2)^2\gamma_5,
\ee
and this will soften the singularity of $L$ compared to eq.~(\ref{finalFermi}).

\begin{figure}[ht]
\begin{center}
\includegraphics[height=19mm]{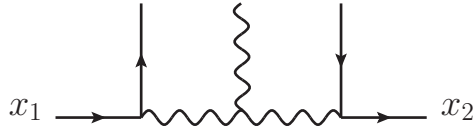}
\end{center}
\caption{Potential contribution of one three-gluon vertex. Such terms are subleading
due to the details of the momentum  of vertices and propagators.\label{fermions_3g}}
\end{figure}

The other potential contributions arise from the gluon self-interactions; an example is shown
in fig.~(\ref{fermions_3g}). It is not difficult to see that, in a general renormalizable gauge,
their momentum dependence leads, after Fourier transform,  to factors of $|x_{12}|^2$
in the numerator which will soften the singularity in the null separation limit and render
these diagrams subleading.

We can therefore conclude that
\be
\lim_{x_{i,i+1}^2\to 0}
\frac{\langle {\cal O}(x_1)\dots {\cal O}(x_n)\rangle}{\langle {\cal O}(x_1)\dots {\cal O}(x_n)\rangle_0}
=
\frac{\langle W_n\rangle_{\text{adj}}}{\langle W_n\rangle_{0,\text{adj}}}
\ee
for 2-field operators ${\cal O}$ constructed from scalars and fermions.

The arguments in \S~\ref{w_insertion} carry over unmodified to the case when one or several
additional operators
are added to such correlation functions and kept at generic positions relative to the operators which
become null-separated; we will not repeat them here. Similarly to the case of scalar operators,
we find that the
$n$-partial null limit, $|x_{i,i+1}|^2\rightarrow 0$ with $i=1,\dots,n$ and $n+1\equiv 1$,
of the connected part of the correlation function of $n+m$ operators is given by
\be
\lim_{x_{ii+1}^2\to 0}
\frac{\langle {\cal O}(x_1)\cdots {\cal O}(x_n){\cal O}(a_1)\cdots{\cal O}(a_m)\rangle^{\text{conn}}}
{\langle {\cal O}(x_1)\cdots {\cal O}(x_n)\rangle}
=
\frac{\langle W_n\,{\cal O}(a_1)\cdots{\cal O}(a_m)\rangle^{\text{conn}}_{\text{adj}}}
{\langle W_n\rangle_{\text{adj}}}  \ .
\ee

\subsection{Gauge fields}

The Feynman diagrams contributing to correlation functions of two-field operators containing
gauge field strengths do not follow the classification we have been using; this is because there
are at least two operators between which there is no direct R-charge flow ({\it i.e.} if a flow exists at
all it must pass through other operators). We will see, however, that unlike previous situation
when absence of a flow led to absence of a sufficiently strong singularity in the null-separation
limit, here a singularity does occur. The main difference compared to previous discussions relates
to the existence of a derivative in the field strength acting along the sequence of propagators not
carrying R-charge. As we will see, this will lead to the desired singularity in the limit in which
a sequence of gluon propagators and three-point vertices stretch between two null-separated points.

Let us illustrate the mechanism at work by analyzing the simplest diagram -- a single gluon
propagator between two operators containing field strengths at points $x_1$ and $x_2$.

\be
L(x_1, x_2)_{\mu\nu}^{\alpha\beta}
=\left(-i\frac{\partial}{\partial x_1^\mu}\eta_{\nu\kappa}+i\frac{\partial}{\partial x_1^\nu}\eta_{\mu\kappa}\right)\eta^{\kappa\lambda}\left(-i\frac{\partial}{\partial x_2^\alpha}\eta_{\beta\lambda}+i\frac{\partial}{\partial x_2^\beta}\eta_{\alpha\lambda}\right)\int d^dp_1\frac{e^{ip_1\cdot(x_1-x_2)}}{p_1^2+i0}.
\ee
The $p_1$ integral is exactly the same as that of a scalar propagator.
Introducing a Schwinger-parameter and carrying out the momentum integral leads again to
the position space scalar propagator, except that now there are two additional derivatives
acting on it.
Each derivative yields a factor of $\zeta^{-1}$, leading to
\be
L(x_1, x_2)&=&{}\frac{1}{4}
\big((x_1-x_2)_\mu\eta_{\nu\kappa}-(x_1-x_2)_\nu\eta_{\mu\kappa}\big)\eta^{\kappa\lambda}
\big((x_2-x_1)_\alpha\eta_{\beta\lambda}-(x_2-x_1)_\beta\eta_{\alpha\lambda}\big)
\nonumber\\
&&\times\int_0^\infty\frac{d\zeta}{\zeta^2}
e^{-i\frac{(x_1-x_2)^2}{4\zeta}-0/\zeta} \ .
\label{gluon_prop}
\ee
This is again an integral of the type (\ref{allints}) and, following (\ref{generic_int}), it will be proportional
to $(|x_1-x_2|^2)^{-3+\epsilon}$. The singularity, stronger than that of a propagator between two scalar
operators, is a reflection of the dimensional analysis.

It is not difficult to deduce the properties of a sequence of gluon propagators and three-gluon vertices.
As in the case of scalar propagators, each gluon propagator brings one positive power of $\zeta$. To
compensate for it and end up with a singularity as strong as (\ref{gluon_prop})  the null limit, each
derivative in the 3-gluon vertex must yield a factor of $\zeta^{-1}$, in close analogy with the derivative
in the scalar-gluon vertex -- see \S~\ref{sgv}. Each such derivative also generates a factor of
$(x_1-x_2)^{\mu_i}$;
due to the antisymmetry of the field strength contributions (cf. eq.~(\ref{gluon_prop})) these factors must
carry the Lorentz indices of the out-going gluons. The details are essentially identical to those in
sec.~\ref{sgv} and we will not repeat them here. The conclusion is that, up to a factor of
$(|x_1-x_2|^2)^{-3+\epsilon}$, the sequence of gluon propagators and three-point vertices between
two null-separated points reduces in the null limit to a Wilson line with as many gluon vertices
as the original number of vertices:
\be
\nonumber
L(x_1,x_2,k)_{\mu\nu}^{\alpha\beta}{}_{\mu_1\dots\mu_n}
&\propto&
\big((x_1-x_2)_\mu\eta_{\nu\kappa}-(x_1-x_2)_\nu\eta_{\mu\kappa}\big)
\eta^{\kappa\lambda}
\big((x_2-x_1)_\alpha\eta_{\beta\lambda}-(x_2-x_1)_\beta\eta_{\alpha\lambda}\big)
\\
&&\times\frac{1}{(|x_1-x_2|^2)^{3-\epsilon}}\prod_{i=1}^n\;i(x_1-x_2)_{\mu_i}\;
\int_{t_{i-1}}^{1} dt_i\;e^{-i\sum_ik_i\cdot (x_2t_i-x_1(1-t_i))} \ ,
\ee
where the field strengths in the two operators at positions $x_1$ and $x_2$ carry indices $(\mu,\nu)$
and $(\alpha,\beta)$, respectively.

Other diagrams, in which the sequence of gluons is interrupted by scalars, fermions or ghosts will have
subleading contributions in the null limit.
Similarly, four-point Lagrangian vertices as well as the commutator terms
in the field strength will drop out for the same reason.

Thus, similarly to the correlation function of operators constructed out of scalars and fermions,
correlation functions of operators
containing field strengths
reduce to the expectation value of a null polygonal Wilson loop with cusps at the positions of the
original operators. As in the case of correlation functions of operators without gauge fields, inclusion of one or more operators at generic positions relative to the null separation points is straightforward with the same result quoted in eqs.~(\ref{ABT}) and (\ref{WW}).



\end{document}